\documentclass[]{emulateapj}
\usepackage{graphicx,latexsym,natbib}

\begin{document}

\title{Are spectral and timing correlations similar in different spectral states in black hole X-ray binaries?}
\author{M. Kalamkar\altaffilmark{1}, M.T. Reynolds\altaffilmark{2}, M. van der Klis\altaffilmark{1}, D. Altamirano\altaffilmark{3}, J.M. Miller\altaffilmark{2}}
\altaffiltext{1}{Astronomical Institute, ``Anton Pannekoek'', University of Amsterdam, Science Park 904, 1098 XH, Amsterdam, The Netherlands}
\altaffiltext{2}{Department of Astronomy, University of Michigan, 1085 S. University Ave., Ann Arbor, MI 48109, USA}
\altaffiltext{3}{School of Physics \& Astronomy, University of Southampton, Southampton, Hampshire SO17 1BJ, UK}
\email{maithili@oa-roma.inaf.it}
\begin{abstract}
\noindent We study the outbursts of the black hole X-ray binaries MAXI J1659--152, SWIFT J1753.5--0127 and GX 339--4 with the \textit{Swift} X-ray Telescope. The bandpass of the X-ray Telescope has access to emission from both components of the accretion flow: the accretion disk and the corona/hot flow. This allows a correlated spectral and variability study, with variability from both components of the accretion flow. We present for the first time, a combined study of the evolution of spectral parameters (disk temperature and radius) and timing parameters (frequency and strength) of \textit{all} power spectral components in different spectral states. Comparison of the correlations in different spectral states shows that the frequency and strength of the power spectral components exhibit dependencies on the disk temperature that are different in the (low-)hard and the hard-intermediate states; most of these correlations that are clearly observed in the hard-intermediate state (in MAXI J1659--152 and GX 339--4) are not seen in the (low-)hard state (in GX 339--4 and SWIFT J1753.5--0127). Also, the responses of the individual frequency components to changes in the disk temperature are markedly different from one component to the next. Hence, the spectral-timing evolution cannot be explained by a single correlation that spans both these spectral states. We discuss our findings in the context of the existing models proposed to explain the origin of variability.\\
\end{abstract}

\section{Introduction}\label{intro}
\noindent Accretion in black hole X-ray binaries (BHBs) can be studied by tracing the evolution of their X-ray spectral and timing properties during outbursts. Decades of studies show evidence of a two-component structure of the accretion flow: a geometrically thick optically thin plasma (which we refer to as the hot flow), and, a geometrically thin optically thick accretion disk. The interplay between these two components leads to dramatic changes in the spectral and variability properties during an outburst. The evolution of an outburst can be studied in terms of different spectral `states'. We first outline the phenomenological behavior of a BHB in a typical outburst \citep[see][for detailed reviews]{belloni2005, homan2005, remillard2006}, followed by a discussion of our understanding of the driving physical processes and the current challenges.\\
\\
The states can be broadly classified as hard and soft states, based on the spectral component that dominates the emission. In the low-intensity hard state (LHS), the energy spectra are dominated by hard emission ($\gtrsim$ 2 keV) modelled by a power-law (index $\lesssim$ 1.8) with a cut-off at few tens of keV.  As the outburst progresses, the intensity increases till the source reaches `intermediate' states (IMSs), divided into hard IMS (HIMS) and soft IMS (SIMS). The energy spectrum gradually `softens' at somewhat constant intensities due to increasing contribution from the soft component, which is modelled by a black-body. Multiple transitions between the HIMS and SIMS are observed before the source makes a transition to the softest state - the high intensity soft state (HSS).  The soft state energy spectra are dominated by the black-body component ($\lesssim$ 2 keV) and a softer and weaker power-law (index $\lesssim$ 2.2). At some point, the intensity decreases and the source goes through the IMS to the LHS during the decline of the outburst.\\
\\
The variability properties in different states are described as follows: the LHS has strong (tens of percent fractional rms amplitude, henceforth \textit{rms}) variability. The power spectra are characterized by broad band noise components, often accompanied by narrow peaked Quasi Periodic Oscillations (QPO) of type-C (see below). The HIMS and SIMS are marked by strongly different variability properties. The HIMS power spectra show type-C QPOs and broad band noise (weaker than in the LHS), while the SIMS power spectra have weaker variability and are often accompanied by either type-A or type-B QPO. Type-C QPOs are stronger and span a larger range of frequencies compared to the type-A/B QPOs. Type-C QPOs are observed in the HIMS, while type-A/B QPOs are observed in the SIMS \citep[see e.g.,][]{wijnands1999qpos, casella2005}.  In the HSS, the variability is very weak (few percent \textit{rms}).\\
\\
The description above is that of a typical BHB outburst. It should be noted that not all sources exhibit all spectral states; e.g., some sources have hard outbursts without transiting to the HSS. The intensity at which different states are observed varies in different sources; e.g., a source may be the hardest (with properties similar to the `LHS') during the peak of the outburst, i.e. at high intensities, while the soft states are exhibited at relatively lower intensities.    \\
\\
The fundamental physical processes that give rise to the different spectral components are understood relatively well. The soft emission modelled by the black-body component is due to thermal emission from the disk. The hard emission is attributed to Compton up-scattering of photons from the disk in an optically thin hot flow. However, the structure and geometry of the accretion flow are under strong debate; it is unclear if the hard component emitting region is a corona, a hot flow \citep[see e.g.,][]{remillard2006, done2007} and/or the base of the jet \citep{markoff2001}. It is also not established how `truncated' the disk is during different stages of the outburst. Earlier, the disk was believed to be far from the black hole (large truncation radii) in the hard state, reaching close to or at the innermost stable circular orbit (ISCO) in the soft state. Recent studies suggest that the disk is not truncated in the hard state (see e.g., \citealt{done2007} and \citealt{reynolds-swift-bhs} for differing arguments).  \\
\\
The origin of variability is somewhat less understood. Although many models have been proposed, a unanimous picture has not emerged yet. For the broad band variability, fluctuations/oscillations/flaring in the accretion flow are considered. Models  for the origin of the QPO mainly fall in two classes:  those associated with a misaligned hot flow  around a spinning black hole \citep{stella1998, fragile2007}  and those associated with oscillation modes in the accretion flow \citep{titarchuk1999, wagoner2001, shaposhnikov-qpo-2012}. Many models however focus mostly on origin of the QPO frequency. Few models which attempt to jointly explain the broad band variability and the type-C QPO are discussed below. \\
\\
\cite{zycki-var-model-2003} attributed broad band variability to multiple active regions/perturbations moving radially towards the central black hole. The QPOs are generated by modulation of one or more parameters of a Comptonization spectrum. In the framework of \cite{cabanac2010}, the variability is associated with an oscillating hot optically thin corona. The oscillations are due to a magneto-acoustic wave that propagates through the corona and modulates the emergent Comptonized emission of soft photons. The power spectrum of the emergent X-ray emission is similar to the band limited noise accompanied by type-C QPO. In another model, fluctuations in the mass accretion rate that propagate through different regions of the accretion flow modulate the emission giving rise to the broad band variability \citep{lyub1997, uttley2005, ingram-prop-fluc-2013}. \cite{ingram-lt-2011} associated the frequencies of different variability components with different radii in the hot flow. The QPO is attributed to the Lense-Thirring precession of the hot flow itself.  Recent studies suggest that part of the broad band variability can also originate in the disk; with the low frequency (few tens of seconds) fluctuations arising intrinsic to the disk \citep{wilkinson, kalamkar1753.5}. \\ 
\\
Detailed spectral studies have been performed using data from various X-ray missions such as EXOSAT, Ginga, Rossi X-ray Timing explorer \citep[RXTE,][]{rxte}, \textit{Swift} \citep{swift} and \textit{XMM-Newton}. Variability studies have been extensively performed with RXTE in the past few years. A study with a combined spectral and timing approach, which covers the emission from both the components of the accretion flow, can help advance our understanding of these systems. Earlier efforts in this direction either mostly reported the QPO, or were performed with RXTE data \citep[see e.g.,][]{grinberg2014-cygx1} which has access only to the hard band. Recent works on disk variability discussed above highlight the need and importance of studying variability in the soft band, which can be accessed with \textit{XMM-Newton} and \textit{Swift}.\\
\\
\textit{Swift} has observed the outburst evolution of many BHBs \citep[see e.g.,][]{reynolds-swift-bhs}.We selected the BHBs MAXI J1659--152, SWIFT J1753.5--0127  and GX 339--4 for our study as they show different behavioral patterns in their outbursts (see Section \ref{1659}--\ref{1753}). This is a unique study, as we investigate the correlated evolution of the spectral components and \textit{all} the timing components with emission from both the components of the accretion flow.  We study the correlations of spectral and timing parameters in different states along the outburst and examine changes across states. In Section \ref{obs}, we introduce the different systems and their earlier reports in detail, and the observations we use in this work. In Section \ref{data-ana}, the methods employed to reduce and analyze the data are discussed. Section \ref{result} discusses the spectral and timing results, and their correlations are discussed in Section \ref{corr-spec-tim}. We discuss our findings and the implications in Section \ref{discussion}. 
\begin{figure*}
\center
\includegraphics[width=4.5cm,height=7cm,angle=-90]{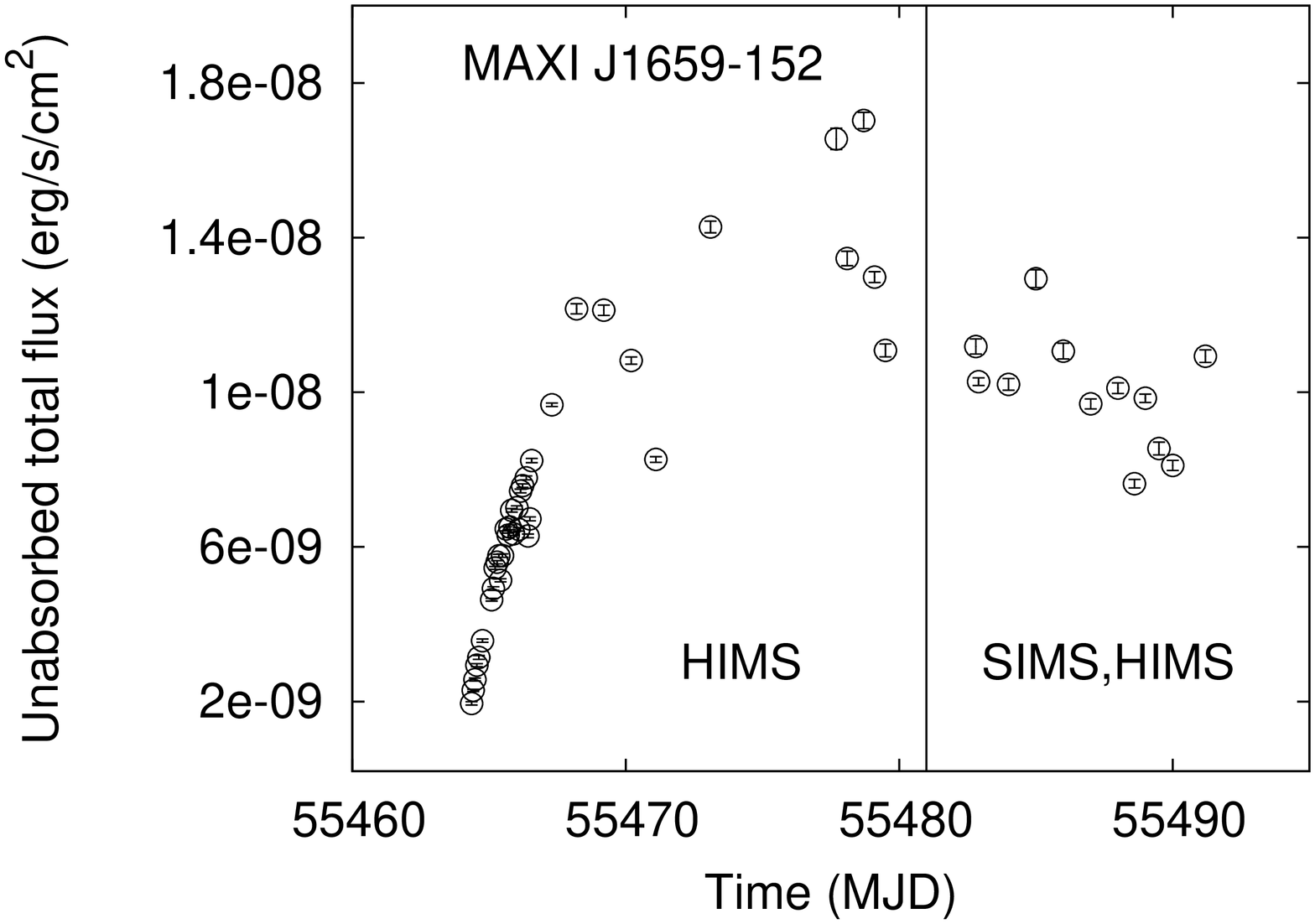}\includegraphics[width=4.5cm,height=5.5cm,angle=-90]{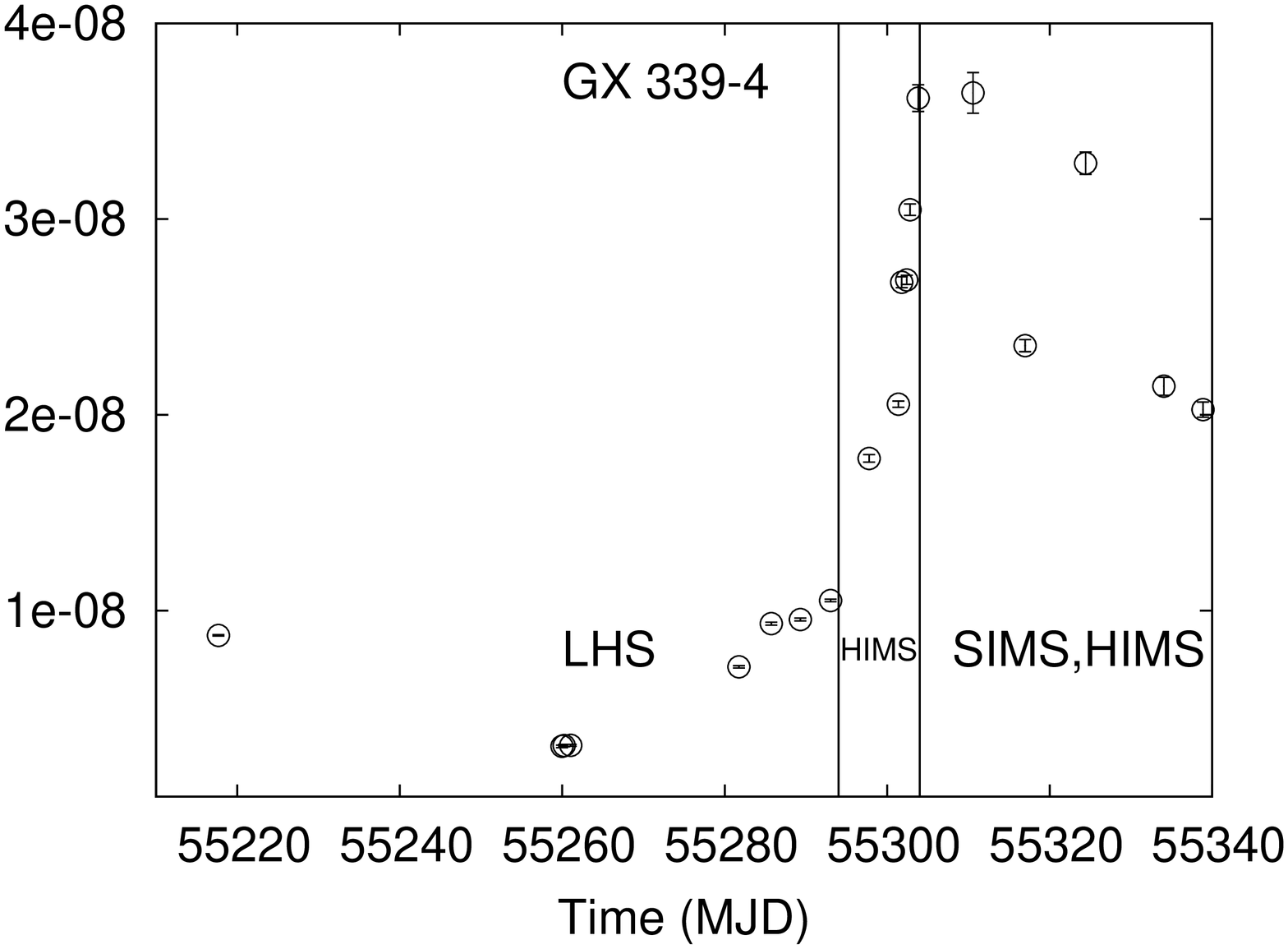}\includegraphics[width=4.5cm,height=5.5cm,angle=-90]{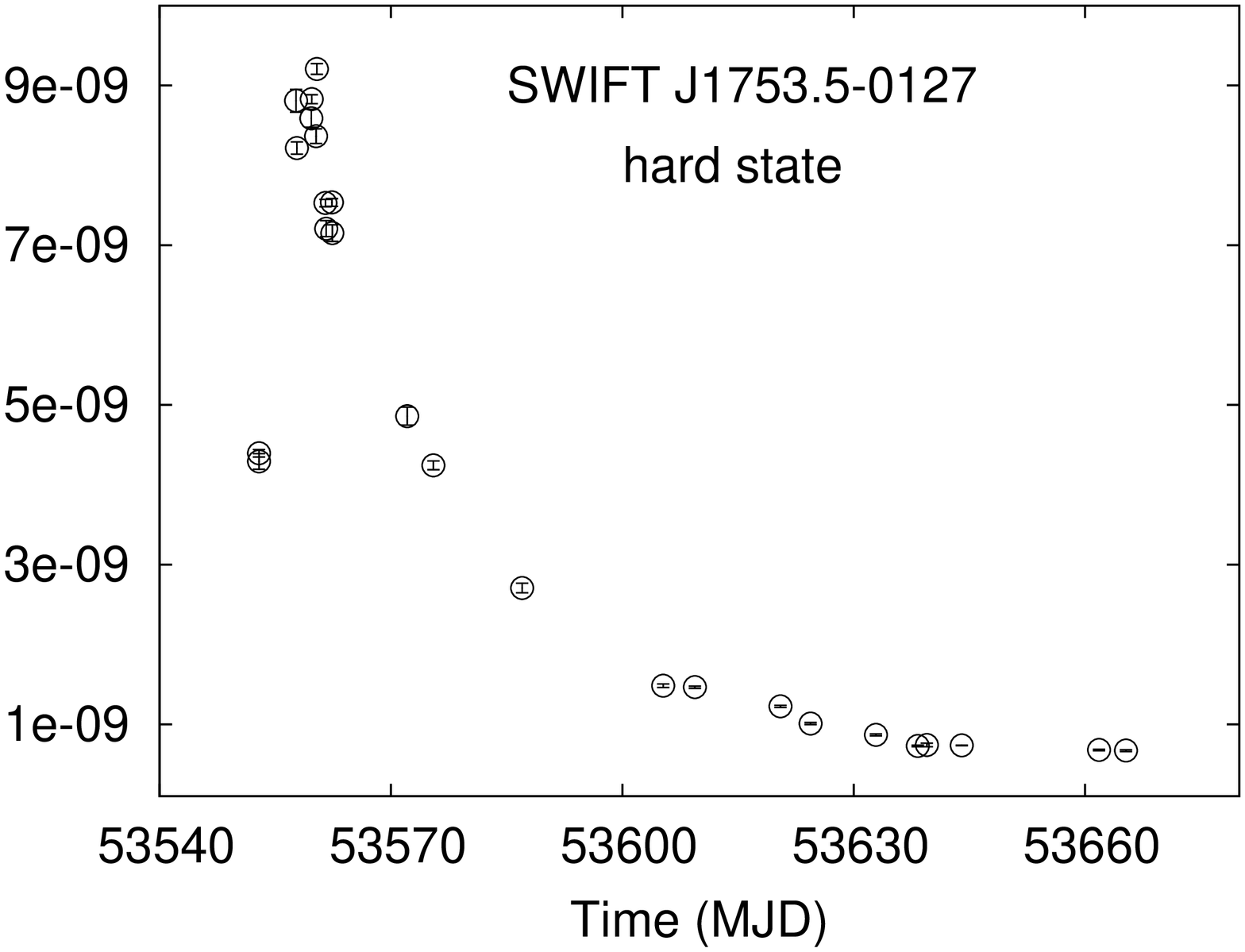}
\caption{Light curves of MAXI J1659--152, GX 339--4 and SWIFT J1753.5--0127. The total unabsorbed flux is in the 0.5--10 keV energy band. The vertical lines in the left and right panels indicate (the first) state transitions; multiple transitions between HIMS and SIMS are observed in both cases, but are not shown here; SWIFT J1753.5--0127 is observed in the hard state with no state transitions in our data.}\label{fig:lc}
\end{figure*}
\section{Observations}\label{obs}
\noindent We study the outburst of the black hole X-ray binaries MAXI J1659--152, SWIFT J1753.5--0127 and GX 339--4. We utilise the observations obtained with the X-ray Telescope \citep[XRT;][] {burrows2005} on board the \textit{Swift} satellite taken in Windowed Timing (WT) mode (in wt2 configuration). Each observation consists of one or more Good Time Intervals (GTIs, times during which XRT was collecting data), which can last up to 2.5 ks. We introduce these sources and describe their observations below.
\begin{figure}
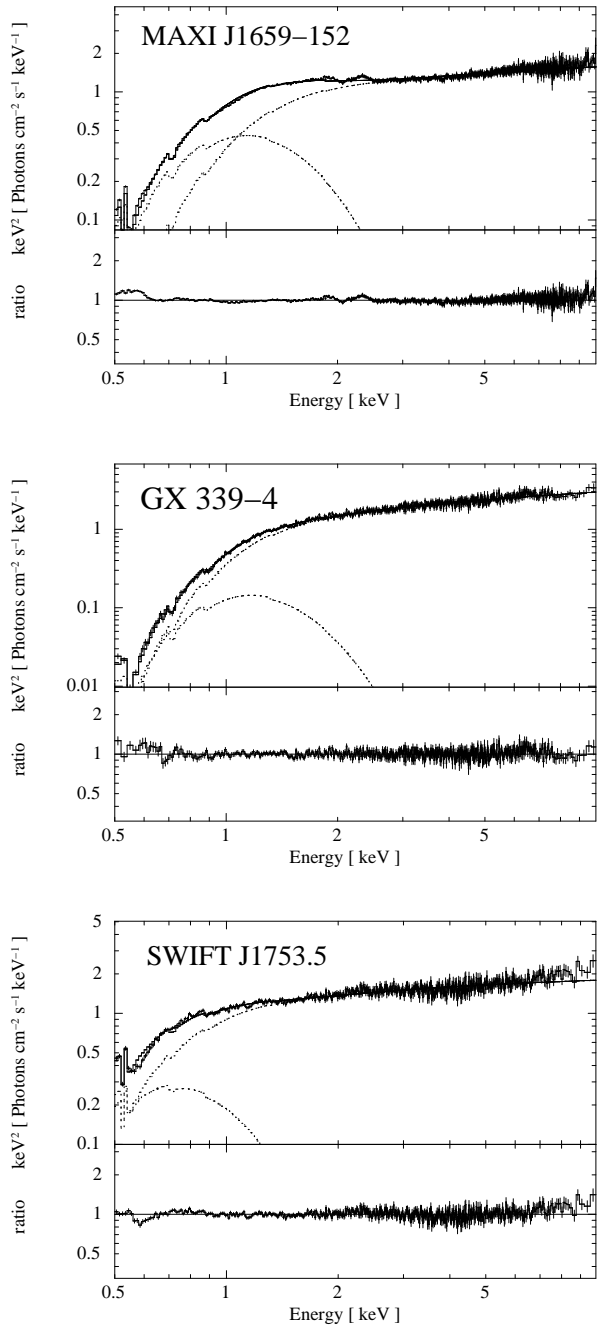

\center
\includegraphics[width=6.05cm,angle=-90]{1659-spectrum.ps}\\
\includegraphics[width=6.05cm,angle=-90]{339-spectrum.ps}\\
\includegraphics[width=6.05cm,angle=-90]{1753-spectrum.ps}
\caption{Representative energy spectra of MAXI J1659--152 (MJD 55466, observation 00434928003), GX 339--4 (MJD 55289, observation 00030943013) and SWIFT J1753.5--0127 (MJD 53562, observation 00030090015) and the residuals (in the respective bottom panels) in the 0.5--10 keV energy band. The spectra are fit with the \texttt{diskbb+comptt} model. See Section \ref{spec-evol} for details. }\label{fig:ener-spec}
\end{figure}
\subsection{MAXI J1659--152}\label{1659}
\noindent MAXI J1659--152 (henceforth J1659) is a BHB suggested to have the shortest known orbital period \citep[2.41 hours;][]{kuulkers1659}. Various reports estimate a distance and the mass of the black hole in the range of  4--8.6 kpc and 2.2--20 Solar masses, respectively. During its (only) outburst in 2010, the X-ray spectral and timing behavior of the source observed with RXTE and \textit{Swift} was similar to that of other BHBs \citep[see][for earlier reports]{kalamkar1659, yamaoka1659, teo16592011, kennea1659, wenfei1659, kuulkers1659, kalamkar1659-swift}. We present the results of 38 observations obtained between September 25, 2010 (MJD 55464) and October 22, 2010 (MJD 55491). The source was in the LHS when it was first observed with the Monitor of All-sky X-ray Image \citep[MAXI;][]{maxi} on MJD 55460.5 \citep{kalamkar1659}. The source had evolved to the HIMS  when \textit{Swift} and RXTE started observing it on MJD 55464 and MJD 55467, respectively. Multiple state transitions to the SIMS have been reported \citep[see][]{kalamkar1659}. The XRT observations ended after the source made the second transition to the SIMS. As noted by \cite{kalamkar1659}, the source did not make a transition to the HSS before returning to the HIMS.
\subsection{GX 339--4}\label{339}
\noindent GX 339--4 (henceforth GX-339) is a BHB located at a distance of $>$ 6 kpc \citep{hynes-339-dist} with a mass function of 5.8 Solar masses and an orbital period of 1.75 days \citep{hynes339}. GX-339 underwent several outbursts and has been reported to exhibit all canonical states observed in BHBs reported with RXTE \citep[see e.g.,][]{zdz-339, motta339}. We present the results of 24 observations of only the outburst in 2010 from January 21 to June 5  (MJD 55217--55352), during which we observe strong variability (QPOs and broad band noise) with \textit{Swift}. The source was in the LHS till MJD 55294 and in the HIMS from MJD 55296--55303 \citep{nandi2012-339, yan2012-339, debnath2010-339}. The source then exhibited state transitions between the HIMS and the SIMS \citep{motta339}, which were covered with XRT. It was reported from the RXTE data that the source then made a transition to the HSS and back to the LHS through HIMS, exhibiting a typical outburst \citep{dincer2012-339, cadolle2011-339}.
\begin{figure}
\center
\includegraphics[width=6cm,angle=-90]{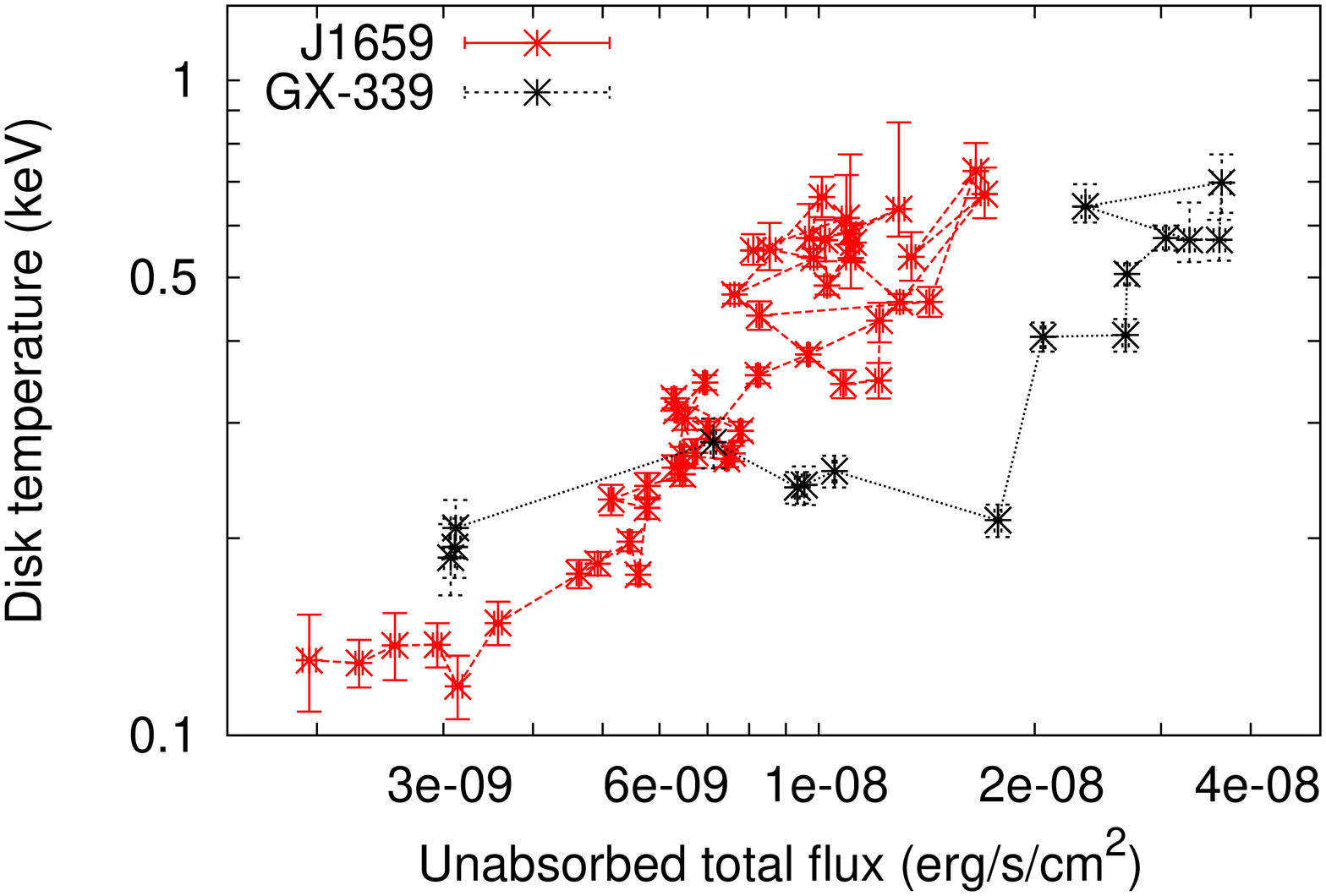}
\includegraphics[width=6cm,height=8.75cm,angle=-90]{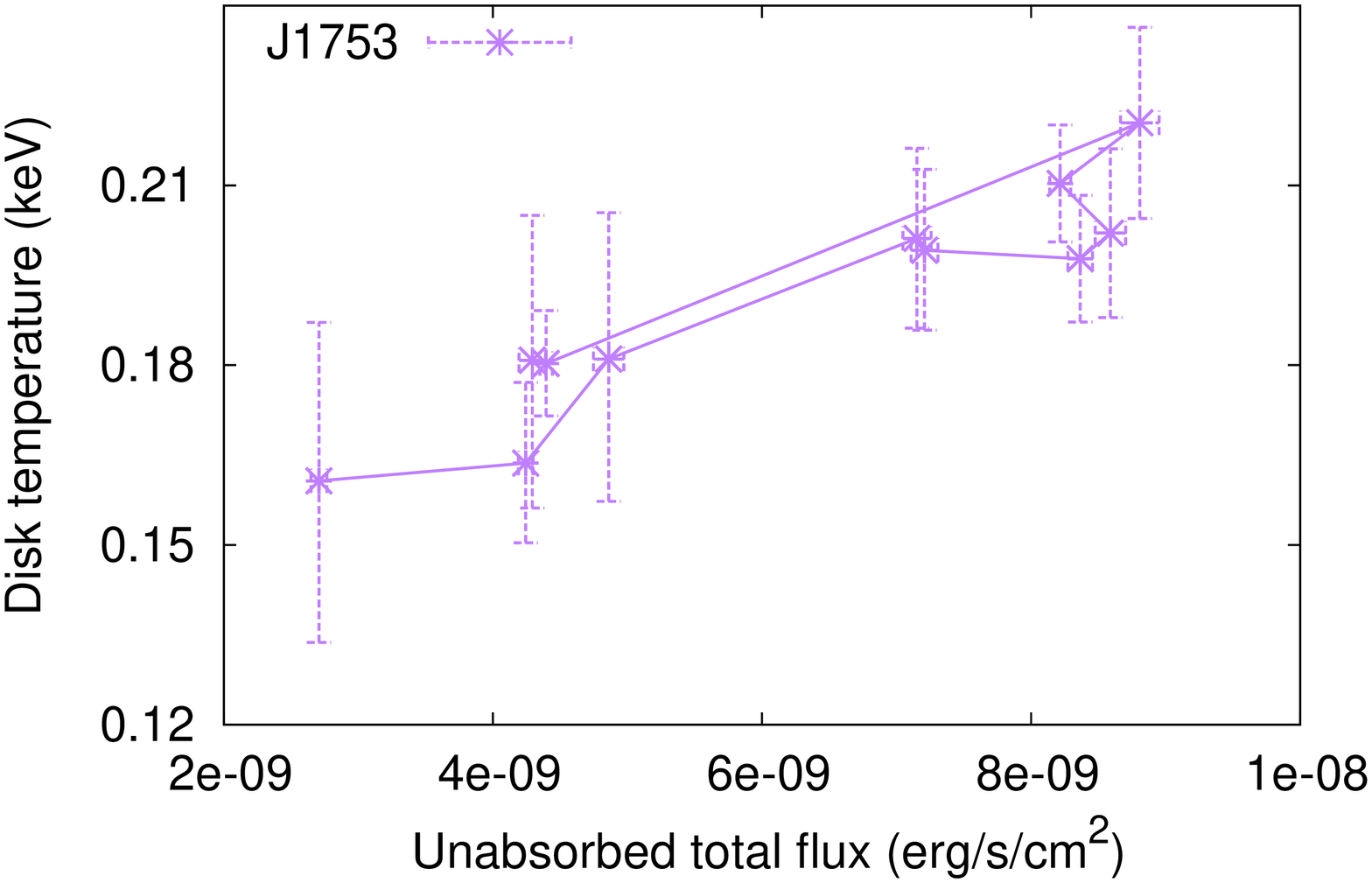}
\caption{Evolution of the disk temperature as a function of the total unabsorbed flux in the 0.5-10 keV energy range. Note the different range of temperatures and fluxes spanned in both the panels.  All detections of the disk shown here are above 5 $\sigma$.}\label{fig:diskt-flux}
\end{figure}
\begin{figure*}
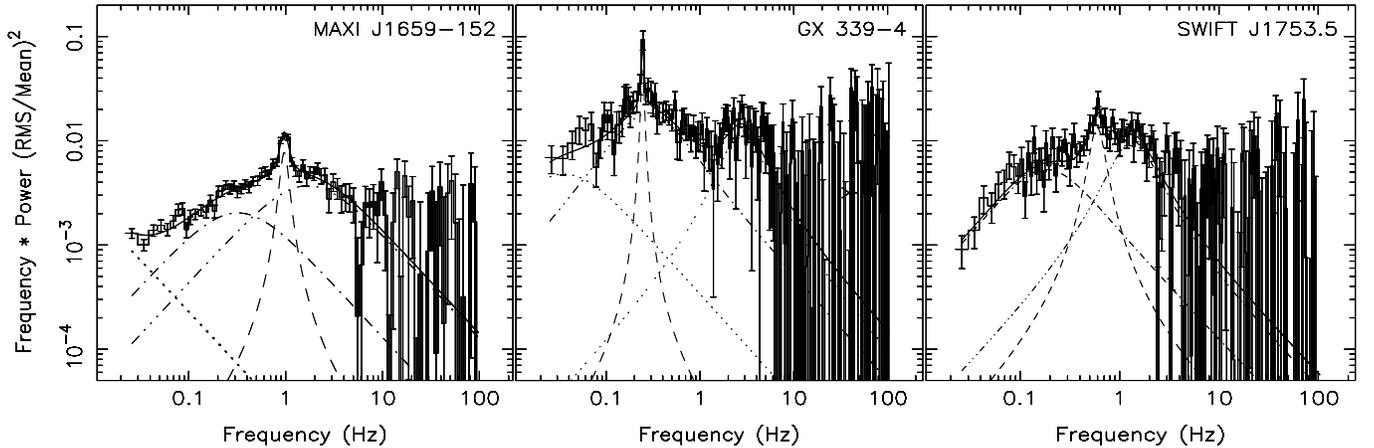

\center
\includegraphics[width=6.75cm,height=5.89cm]{1659.ps}\includegraphics[width=5.45cm,height=5.89cm]{339.ps}\includegraphics[width=5.75cm,height=5.89cm]{1753.ps}
\caption{Representative power spectra of MAXI J1659--152 (MJD 55466, observation 00434928003), GX 339--4 (MJD 55289, observation 00030943013) and SWIFT J1753.5--0127 (MJD 53562, observation 00030090015) in the 0.5--10 keV energy band. The best-fit model using multiple Lorentzians is shown. See Section \ref{tim-evol} for their identification. }\label{fig:pds}
\end{figure*}
\subsection{SWIFT J1753.5--0127}\label{1753}
\noindent SWIFT J1753.5--0127 (henceforth J1753) is the BHB with the second shortest known orbital period \citep[3.2 hours;][]{zurita1753}, and is a quasi-persistent system. The estimated distance and the black hole mass are between 4--8 kpc and 4--16 Solar masses, respectively \citep{cadolle, zurita1753}. It went into an outburst in 2005 and has been active since then. Spectral and timing studies revealed that the source remained in the (low) hard state during the peak and decay of the outburst in 2005 \citep{cadolle, zhang2007, ramadevi, chiang2010}. It made a transition to the HIMS after four years, but did not make a transition to the SIMS or the HSS \citep{soleri1753}. Hence, J1753 did not show the typical evolution through the different states of an outburst seen in other BHBs. Spectral analyses with \textit{XMM-Newton} and RXTE suggested the presence of a cool disk at or close to the ISCO in the hard state \citep{miller2006, chiang2010}. We present the results of only the 29 bright observations obtained in 2005 between July 1 and October 22 (MJD 53552--55337) during the hard state, where we observe QPOs and broad band noise components. The spectral and timing evolution of the source have been reported by \cite{reynolds-swift-bhs} and \cite{kalamkar1753.5}, respectively.\\
\\
The three sources discussed above show varied differences in their outburst properties. J1659 and GX-339 exhibit typical outburst behavior, except J1659 did not make a transition to the HSS. J1753 is a quasi persistent system in outburst since 2005. It remained in the hard state during the rise and decay of the outburst in 2005. Despite varied outburst behavior, some of the timing properties are quite similar in all the systems, e.g., the presence of type-C QPO. The spectral properties show the presence of disk emission in some of the hard state observations. All these properties of the three sources make them suitable for our correlated spectral-timing study.  
\section{Reduction \& Analysis Procedure}\label{data-ana}
\subsection{Spectral Analysis}\label{spec-ana}
\noindent All observations were reprocessed using \texttt{xrtpipeline} and the latest \textit{Swift} \textsc{caldb} files. Source and background spectra in addition to the relevant response files and exposure maps were created as outlined in \citet{reynolds-swift-bhs}, where we consider valid events in the 0.5--10 keV energy range. Data reduction and analysis were performed within the \textsc{heasoft 6.13} environment containing \textsc{xspec 12.8.0j} \citep{arnaud1996}. All spectra have their energy channels grouped into bins to contain a minimum of 20 counts per bin. Errors are calculated via the \texttt{error} command and are equivalent to the 90\% confidence interval. We report the average energy spectrum per observation, except for the observations of J1659 between MJD 55464--55466, where the energy spectrum is reported for each GTI (see below). \\
\\
We follow the procedure outlined in \citet{reynolds-swift-bhs} to analyze the energy spectra. Here we present the results of the model \texttt{phabs(diskbb+comptt)}. The comptonization model (\texttt{comptt}) rolls over at low energies and we fix the \texttt{comptt} input seed photon temperature to that of the disk. Accounting for this is of the utmost importance for detectors with a low energy cutoff such as \textit{Swift} XRT. \\
\\
Recently \citet{miller2009} have shown that there is a lack of significant local absorption in LMXBs. Hence, the column density was fixed at a value consistent with the literature. Absorption by intervening neutral hydrogen was modeled via \texttt{phabs}, where the abundances and cross-sections assumed are \texttt{bcmc} \citep{bcmc92} and \texttt{aspl} \citep{aspl09} respectively. For completeness, fits were also carried out with the \texttt{tbnew} absorption model \citep{wilms11} and the results are found to be consistent with \texttt{phabs} at all times.
\subsection{Timing Analysis}\label{tim-ana}
\noindent The observations were processed using the standard procedure of \cite{evans2007}. We select only grade 0 events, as these are predominantly single pixel events\footnote{The incident photon may generate a charge cloud spread over multiple pixels. Events spread over multiple pixels lead to drop-off in power at high frequencies in the power spectrum \citep{kalamkar1753.5}}. To obtain the XRT power spectrum (after removal of data affected by pile-up and without background or bad pixel corrections), we use the procedure described by \cite{kalamkar1753.5}. Leahy normalized \citep{leahy1983} power spectra were generated of 115.74-s continuous intervals (or shorter if the Good Time Interval was short). The power spectra were then averaged per GTI for the observations of J1659 between MJD 55464--55466, as the GTIs were a few hundred seconds long, and per observation for the rest of the data. As the time resolution of the WT mode is 1.766 ms, the Nyquist frequency is 283.126 Hz.\\
\\
We analyze the power spectra below 100 Hz. The Poisson level is estimated by averaging the power between 50--100 Hz, where no source variability is observed (we refer the reader to the Appendix in \citealt{kalamkar1753.5} for details). This estimated Poisson level is subtracted from each power spectrum of the GTI/observation, and the power spectrum is expressed in rms normalization \citep{vdk1989}. The power spectra are fit with several Lorentzians in the `$\nu_{max}$' representation \citep{belloni2002}. The fit parameters for each Lorentzian are: the characteristic frequency $\nu _{\rm max} \equiv \nu _{0}\sqrt{1+1/(4Q^{2})}$, the coherence or the quality factor Q$\equiv\nu_{0}$/FWHM, and the integrated power $P$, where $\nu_{0}$ is the centroid frequency and FWHM is the full width at half maximum of the Lorentzian. When the Q $<$ 0.0 in the fit, it is fixed to 0.0; this did not affect the other parameters significantly. All the components reported have a single trial significance of $P/\sigma_{P_-}$ $>$ 3.0, with $\sigma_{P_-}$ the negative error on $P$ calculated using $\Delta\chi^2$ = 1. The  errors on timing parameters reported here are calculated using $\Delta\chi^2$ = 1. We study the power spectra in two energy bands: 0.5--2 keV and 2--10 keV, which we will refer to as soft and hard bands, respectively. 
\section{Results}\label{result}
\subsection{Light curves}\label{lc-evol}
\noindent Figure \ref{fig:lc} shows the light curve of the three sources. The different states in which these sources were observed are indicated. In J1659, the nature of the light curve was reported to be fast rise and exponential decay \citep{kennea1659, yamaoka1659}. The source was first observed when it was already in the HIMS; transitions between  the HIMS and  the SIMS were observed during the later part of the outburst \citep{kalamkar1659}. In J1753, the nature of the light curve was also reported to be fast rise and exponential decay. It should be noted that the source was in the hard state during all these observations and that this is a hard state at `high' intensity, as it is observed during the peak of the outburst \citep{cadolle, ramadevi, soleri1753, zhang2007}.  GX-339 was observed in the LHS,  the HIMS and  the SIMS with several transitions between the HIMS and the  SIMS \citep{motta339}. It should be noted that unlike J1659 and J1753, the light curve of GX-339 was reported to be slow rise slow decay \citep{debnath2010-339}.
\subsection{Spectral evolution}\label{spec-evol}
\noindent Figure \ref{fig:ener-spec} shows representative energy spectra of the three sources. The spectra are in the HIMS  for J1659, in the hard state for J1753 and in the LHS for GX-339. They all show the presence of the soft disk component and hard power-law like emission in these observations modelled with \texttt{diskbb+comptt}. The disk is significantly detected in many observations in these three sources (see below).\\ 
\\
Figure \ref{fig:diskt-flux} shows the evolution of the disk temperature as a function of the total unabsorbed flux.  In J1659, the disk is detected over the entire period of observations. The temperature initially stays somewhat constant, followed by an increase in a correlated fashion with the flux. The correlation spans the HIMS and the SIMS, although a large scatter is seen at higher disk temperatures. In GX-339, the disk is not significantly detected in the first observation (and hence not shown in Figure \ref{fig:diskt-flux}), but it is detected in all subsequent observations. The temperature does not show a large change during the LHS \citep[as also reported by][]{cadolle2011-339}. The temperature begins to increase after the source enters the HIMS, (flux $>$ 1.7e-08  erg/s/$cm^{2}$). A scatter is seen above a disk temperature of 0.5 keV, corresponding to the time when the source exhibits transitions between the HIMS and the SIMS. In J1753, the disk is detected during the rise, peak and decay of the outburst till MJD 53587 (see Figure \ref{fig:lc}). In Figure \ref{fig:diskt-flux}, it appears that the temperature increases till the flux reaches its maximum, followed by a decrease in temperature during the flux decay. However, as the errors on the disk temperature are large, this cannot be said conclusively. Independent of the large errors, we observe that both the disk temperature  and the flux vary over a limited range in J1753; J1659 and GX-339 span a larger range of fluxes as well as disk temperatures.\\
\\
The important spectral parameters that characterise the spectral model are the disk temperature (which in our model is equal to the input seed photon temperature) and the plasma optical depth. As the XRT CCD is only sensitive to X-ray emission in the energy range below 10 keV, we are unable to independently constrain the electron temperature, which requires detection of the spectral cut-off typically present at energies in excess of 10 keV. For this reason, the electron temperature was fixed at 50 keV in all fits. This has the effect of producing power-law like high energy emission in the XRT bandpass. The optical depth and electron temperature are known to be somewhat degenerate in the \texttt{comptt} model, hence, we are unable to uniquely constrain the absolute value of the optical depth. For this reason, although we fit the spectra with \texttt{diskbb+comptt} accounting for both the components of the accretion flow, here we present the evolution of, and correlations with only the disk parameters (temperature and radius). We emphasise that our choice of electron temperature does not affect our measurement of the disk parameters. 
\subsection{Timing evolution}\label{tim-evol}
\noindent Figure \ref{fig:pds} shows representative XRT power spectra from the sources in the 0.5--10 keV energy band, using the same observations as in Figure \ref{fig:ener-spec}. The components in each power spectrum are identified as follows: The power spectrum of J1659 (HIMS) shows four components which, in the order of increasing frequency, are the low frequency noise (lfn), the `break' component, the QPO and the broad band noise underlying the QPO, referred to as the `hump', as identified in \cite{kalamkar1659} with the RXTE data and in \cite{kalamkar1659-swift} with the \textit{Swift} data.  In GX-339 (LHS),  based on the frequency and \textit{rms} evolution properties we observe (see below), we identify the components as the lfn, the QPO \citep{motta339}, the hump and an additional component seen around 3 Hz. A similar component was reported during the rise of the 2002/2003 outburst \citep{belloni2005}. As we detect this component only in GX-339, we do not study it further. The break is not detected during these 2010 XRT observations. The break was reported in very few observations during the 2002/2003 observations of GX-339 \citep{belloni2005}. The power spectrum of J1753 is of an observation from the peak of the outburst, during which the source was in the `hard' state. Except for the lfn (not detected in any observation), the same components are detected in J1753 as described for J1659, as identified in \cite{kalamkar1753.5}. \\

\begin{figure}
\center
\includegraphics[width=12cm,height=10cm,angle=-90]{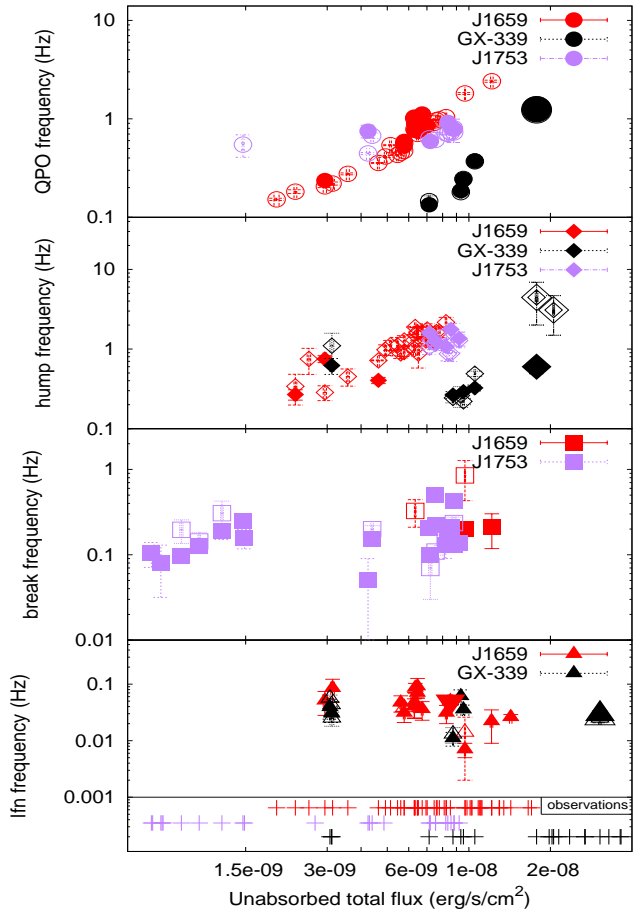}
\caption{Evolution of the frequency of the different variability components as a function of the total unabsorbed flux in the 0.5-10 keV energy range. The bottom panel indicates the coverage of the observations;  the components are not always detected. The states covered are: J1659 - HIMS and SIMS (lfn only - large inverted triangle), GX 339 - LHS (larger sized symbols) and HIMS, and, J1753 - hard state. The filled and open symbols show the frequency in the 0.5--2 keV and 2--10 keV bands, respectively.}\label{fig:freq-flux}
\end{figure}
\begin{table}
\centering
{\renewcommand{\arraystretch}{}
\begin{tabular}{|c|c|c|c|c|}
\hline 
Source & lfn & break & QPO & hump \\
& & & type-C &\\
\hline
J1659 & HIMS, SIMS & HIMS & HIMS & HIMS\\
\hline
J1753 & -- & hard state & hard state & hard state\\
\hline
GX-339 & LHS, HIMS & -- & LHS, HIMS & LHS, HIMS \\
\hline
\end{tabular}}
\caption{The table represents the detections of various variability components in the three sources in different spectral states. See Section \ref{lc-evol} for the discussion on spectral states and Section \ref{tim-evol} for the identification of the components.}\label{tab:det}
\end{table}
\begin{figure}
\center
\includegraphics[width=12cm,height=10cm,angle=-90]{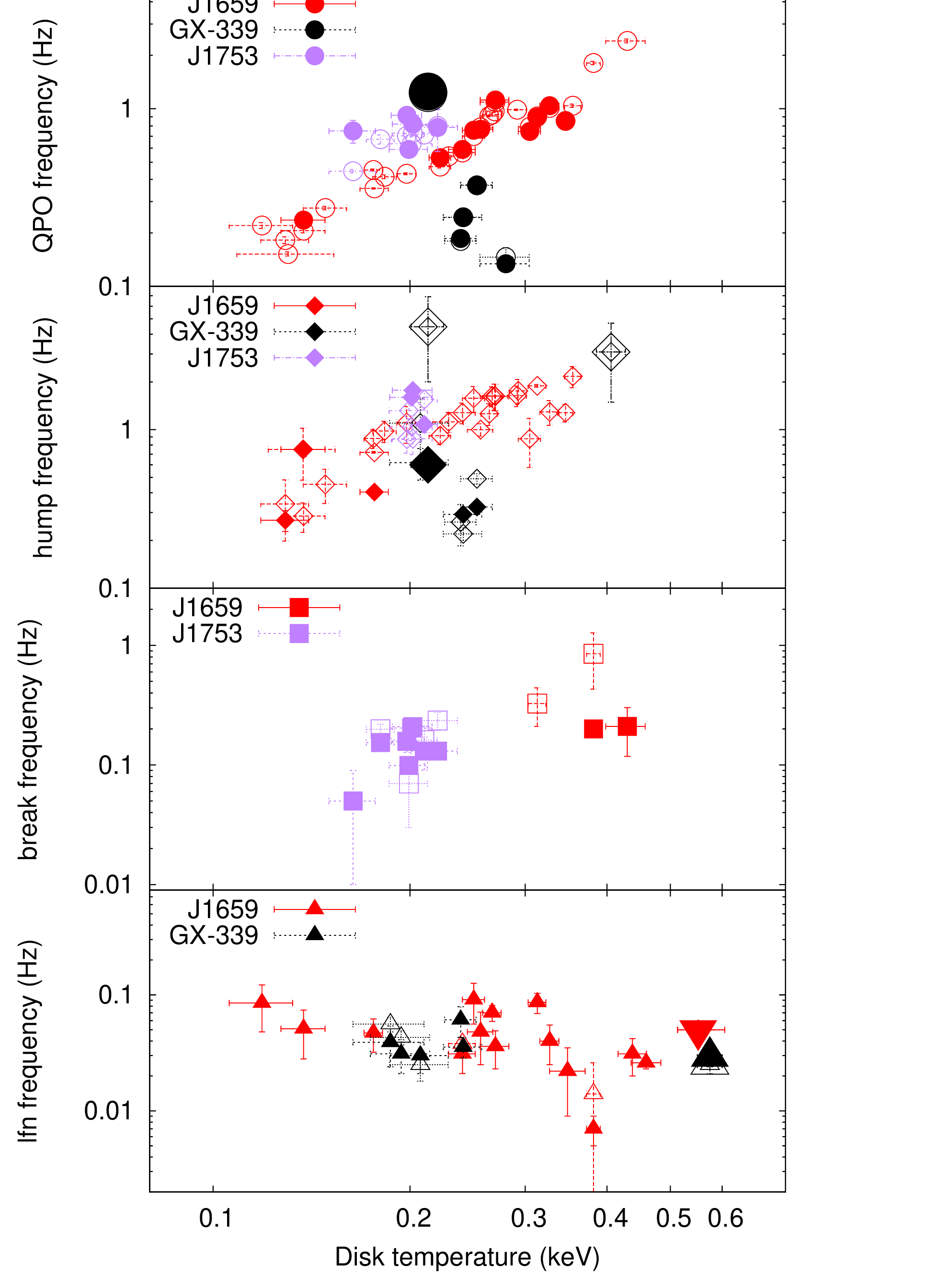}
\caption{Evolution of the frequency of the different variability components as a function of the disk temperature. The states covered are: J1659 - HIMS and SIMS (lfn only - large inverted triangle), GX 339 - LHS (larger sized symbols) and HIMS, and, J1753 - hard state. The filled and open symbols show the frequency in the 0.5--2 keV and 2--10 keV bands, respectively.}\label{fig:freq-diskt}
\end{figure}
\begin{figure}
\center
\includegraphics[width=12cm,height=9.5cm,angle=-90]{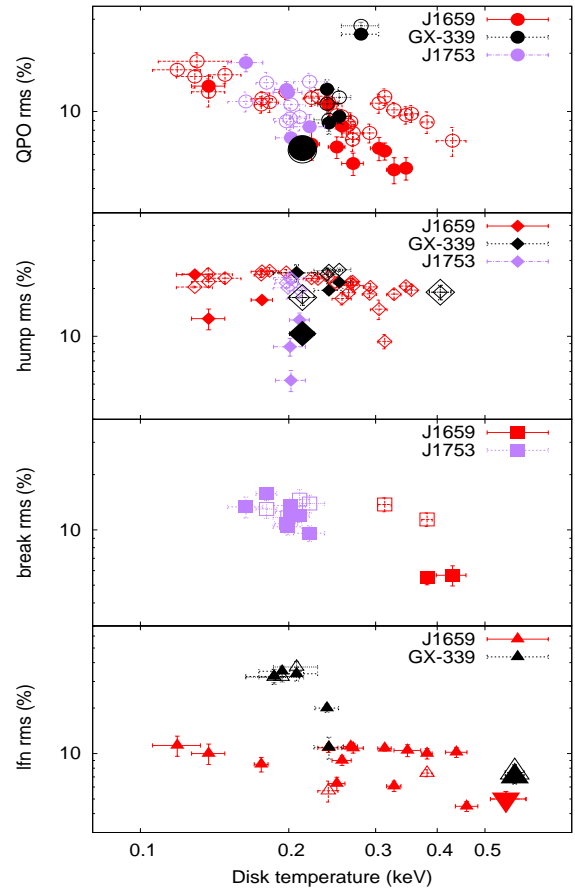}
\caption{Evolution of the fractional rms amplitude of the different variability components as a function of the disk temperature. The states covered are: J1659 - HIMS and SIMS (lfn only - large inverted triangle), GX 339 - LHS (larger sized symbols) and HIMS, and, J1753 - hard state. The filled and open symbols show the fractional rms amplitude in the 0.5--2 keV and 2--10 keV bands, respectively.}\label{fig:rms-diskt}
\end{figure}
\noindent The detections of the various components in different spectral states in the three sources are shown in Table \ref{tab:det}. These components are typical of the LHS and the HIMS, and have been reported in many BHBs \citep[see, e.g.,][]{klis2006}. We detect type C QPO in all the sources; type B and type A QPOs which are typical of the SIMS, are not detected in our data. It should be noted that not all components are seen in each source in each observation. All the components are detected in the hard and soft bands, although not always simultaneously. Interestingly, we detect the QPO and the hump components more often in the hard band, while the break and lfn components are detected more often in the soft band. Figure \ref{fig:freq-flux} shows the frequency evolution of the variability components as a function of the total unabsorbed flux. In BHBs, the component frequencies  generally correlate with flux. We observe that:\\
\\
1) The QPO frequency is strongly correlated with flux in J1659 (HIMS) and GX-339 (LHS and HIMS, see below), but in our data the correlation is not clear in J1753 (hard state). A strong correlation has been reported in J1753 with the RXTE data \citep{zhang2007, ramadevi} during the decay.\\
\\
2) The hump frequency shows a strong correlation with flux for J1659 (HIMS), but is not clearly seen in J1753 (hard state) in our data. In GX-339, the hump frequency does not exhibit a clear correlation with the flux during the observations at low flux, which are in the LHS. It increases sharply only during the two detections in the hard band which correspond to the HIMS; the frequency is higher in the hard band than the soft band. \\
\\
3) The break component in J1659 (HIMS) is detected much later along the outburst than the rest of the components and shows only two detections. The frequency shows an increase with flux only in the hard band. Correlation of the frequency with intensity has been reported with the RXTE data in the 2-60 keV range \citep{kalamkar1659-swift}. The frequency is higher in the hard band than the soft band for the only simultaneous detection. Such an energy dependence of break frequency (frequency increasing with energy) has been reported in this source \citep{kalamkar1659-swift} and also in other sources \citep{belloni1997,kalemci2003-1650}. In J1753 (hard state) during the peak of the outburst,  the break frequency does not show a clear dependence on flux, but during the flux decay (below 1.5e-08 erg/s/$cm^2$) the frequency decreases. The break component is not detected in GX-339. \\
\\
4) The frequency of the lfn component varies in the range 0.01-0.1 Hz with no clear flux dependence over a large range of fluxes in J1659 and GX-339, which in the case of J1659 is across the HIMS and the SIMS (there is one detection of the lfn during the SIMS) and during the LHS and the HIMS in GX-339. The lfn is not detected in J1753.
\section{Spectral--timing correlations}\label{corr-spec-tim}
\noindent In this section, we discuss the correlations of the spectral and timing parameters. Each  source spans a different (but overlapping) range of fluxes and disk temperatures; yet, the variability components are seen over very similar range of frequencies in these sources. Strong variability is observed in the hard states, which becomes weaker as the source evolves to the soft states i.e., as the disk temperature increases. We study how the disk evolution affects the variability in the spectral states of these sources, which exhibit different behavior along the outburst evolution.
\subsection{Frequency vs. Disk temperature}\label{sec:freq-diskt}
\noindent Figure  \ref{fig:freq-diskt} shows how the frequencies of different components vary with the disk temperature. In the canonical BHB outburst behavior, the frequencies of various components increase along with the disk temperature as the source moves to softer states. We observe that:\\
\\
1) The QPO frequency is strongly correlated with the disk temperature only in the case of J1659 (HIMS). The behavior of J1753 and GX-339 can be inferred with additional information from earlier reports. In J1753 (hard state), the QPO frequency has been reported to decrease during the decay \citep{zhang2007, ramadevi}, and we see a possible small decrease in disk temperature in our decay data (although with large errors) suggesting a correlation. In GX-339, the disk temperature increases during the HIMS. We have only one detection of the QPO in the HIMS in our data. \cite{motta339} and \cite{nandi2012-339} reported an increase in QPO frequency during the HIMS, and hence the frequency and the disk temperature are probably correlated. Interestingly, in the LHS we observe that the frequency increases but the disk temperature does not change much (as seen in Figure \ref{fig:diskt-flux}). Hence the frequency-disk temperature correlation is seen in the HIMS, but is not clearly seen in the LHS. This indicates that this correlation has a strong dependence on the spectral state of the source. It is interesting to note that the correlation of the QPO frequency with flux in GX-339 shown in Figure \ref{fig:freq-flux} is smoother and state independent. \\
\\
2) The hump frequency is strongly correlated with the disk temperature only in the case of J1659 (HIMS); the correlation is not seen in J1753 or in GX-339.\\
\\
3) The break and lfn frequency do not show a clear dependence on the disk temperature in any of the sources (except the break component in J1659 which shows a correlation only in the hard band). This behavior of the lfn is similar to what is seen in Figure 4, where no dependence of the frequency on the flux is observed.
\subsection{Fractional rms amplitude vs. Disk temperature}\label{sec:rms-diskt}
\noindent We study the dependence of the \textit{rms} of the various frequency components in the power spectra on the disk temperature. It is commonly seen in BHBs that the \textit{rms} of these components decreases as the total flux increases \citep[see, e.g.,][]{bhb-mcclintock-2006}. The disk temperature and the disk contribution to the flux increase when the source evolves to softer states in a typical outburst. As it has been suggested that different variability components have their origin in different regions of the accretion flow: the disk and/or the hot flow \citep{wilkinson, kalamkar1753.5}, it is worth investigating if the increasing disk temperature has the same, or different effects on the strength of the different components in different states. Figure \ref{fig:rms-diskt} shows the dependence of the \textit{rms} of different power spectral components on the disk temperature. We observe that:\\
\\
1) The \textit{rms} of the QPO decreases with increase in disk temperature in J1659 (HIMS) and J1753 (hard state). The \textit{rms} decreases more steeply in the soft band compared to the hard band. In GX-339, we have only one detection of the QPO in the HIMS and we infer the behavior during the HIMS from an earlier report. \cite{nandi2012-339} reported that the QPO \textit{rms} decreases during the HIMS as the source evolves to the SIMS using RXTE data, during the period where we observe an increase in disk temperature. Hence, the QPO \textit{rms} and disk temperature are probably anti-correlated during the HIMS. However during the LHS, we do not observe the same anti-correlation; the \textit{rms} decreases as the source evolves to the HIMS, but not in an anti-correlated fashion with the disk temperature as seen in Figure \ref{fig:rms-diskt}. As discussed in Section \ref{spec-evol}, the disk temperature does not show large change during the LHS (there appears to be a slight decrease in the disk temperature before it starts increasing in the HIMS).\\
\\
\noindent 2) The \textit{rms} of the hump does not change much over a large range of disk temperature in J1659 (HIMS) and GX-339 (LHS and HIMS). In  J1753, the \textit{rms} is similar in all the hard state observations during which the disk is detected. The component is stronger in the hard band than the soft band in J1659 and J1753, more prominently so in J1753.\\
\\
3) For the break component in J1753 (hard state), we observe an increase in the strength during the decay \citep[also reported in][]{kalamkar1753.5}; as the disk is not significantly detected in these observations, these data points are not present in Figure \ref{fig:rms-diskt}. In J1659 (HIMS) with the RXTE data, \cite{kalamkar1659-swift} reported an increase in \textit{rms} as the source evolves to softer states, during which we observe an increase in disk temperature, indicating a positive correlation. \\
\\
4) The lfn in GX-339 is strongly (50 \% \textit{rms} in the soft band and 41 \% \textit{rms} in the hard band) detected during the first observation in the LHS when the disk is not significantly detected (and hence not seen in the Figure). The \textit{rms} decreases during the rest of the observations in the LHS, falling to values close to the strength of lfn in J1659 during the HIMS. The lfn strength stays fairly stable (with some scatter) across a large range of disk temperature in J1659 (HIMS). The lfn is the only component detected in the SIMS in our data in J1659. Hence, the \textit{rms} of the hump, break and lfn do not show a clear correlation with the disk temperature, such as that seen in the case of the QPO. 
\begin{figure}
\center
\includegraphics[width=6.25cm,angle=-90]{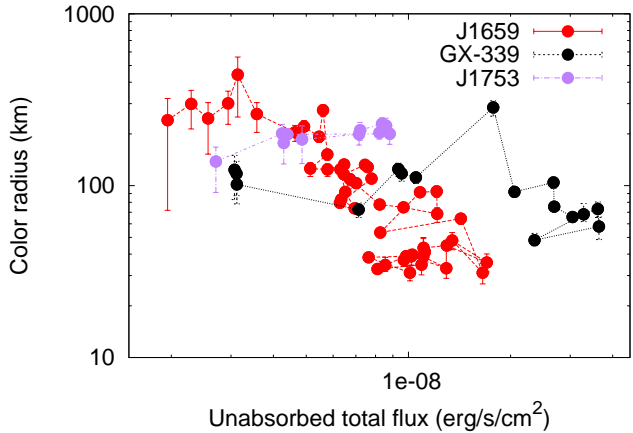}
\caption{Evolution of the color radius of the disk as a function of the total unabsorbed flux in the 0.5-10 keV energy range. We assume a distance of 6 kpc and the inclination angle has been set to 0$^{\circ}$. For an inclination of 60$^{\circ}$, the radii will be larger by a factor of 1.4.}\label{fig:rad-flux}
\end{figure}
\subsection{Inner disk radius evolution}\label{sec:radius-evol}
\noindent The inner radius of the disk (truncation radius) can be estimated from the normalization of the \texttt{diskbb} model, given as r$_{in}$[km]= d$_{10kpc}$*$\sqrt{(normalization/cos \theta)}$, where d is the distance to the source (in kpc) and $\theta$ is the inclination angle. As the distance, the inclination angle and the BH mass are not precisely known for these sources, there are uncertainties in the estimated radius. Also, one should take into account factors such as: spectral hardening \citep{shimura1995}, and the disk temperature not peaking at the inner radius \citep{kubota1998}, which gives the combined corrected radius as {1.18}*r$_{in}$[km]. Hence, it should be noted that these radii are the apparent radii \citep[see, e.g.,][]{reynolds-swift-bhs}. However, the trends in the \textit{changes} in the radius are expected to be robust which is our focus here. \\
\\
Figure \ref{fig:rad-flux} shows the disk color radius as a function of the total unabsorbed flux. The sources show very different behavior. In J1659 (HIMS) the disk initially appears to stay at similar radii, and then decreases i.e., the disk moves inwards, as the source evolves to the soft state. In J1753 (hard state), the radius does not show large changes \citep[also see][]{chiang2010}, as compared to J1659, and shows a possible decrease at low intensity. In GX-339, the radius stays at similar values (close to $\sim$ 100 km) in the LHS. During the first observation in the HIMS ($\sim$ 300 km), the radius increases by a large factor. During the rest of the HIMS observations, the radius  decreases initially and stays at values lower than during the LHS ($\sim$ 100 km). \\

\begin{figure}
\center
\includegraphics[width=6.25cm,angle=-90]{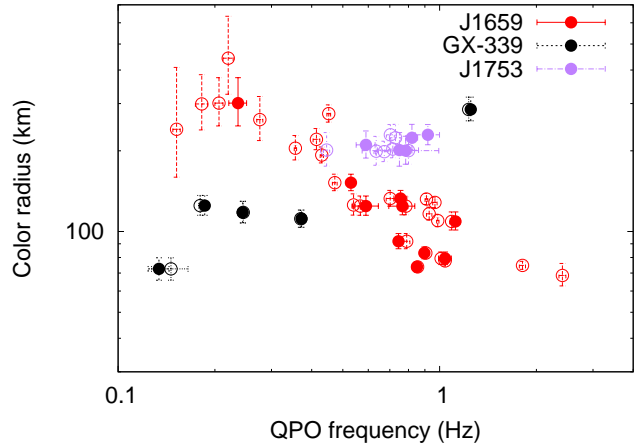}
\caption{Evolution of the color radius of the disk as a function of the frequency of the QPO. We assume a distance of 6 kpc and a black hole mass of 10 Solar masses. The inclination angle has been set to 0$^{\circ}$. }\label{fig:qpofreq-diskrad}
\end{figure}
\noindent The variation of the inner disk radius is often associated with the changes in the frequency of the components in the power spectrum, and an anti-correlation between the frequency and the radius is expected. Figure \ref{fig:qpofreq-diskrad} shows the relation between the color radius and the QPO frequency. In J1659 (HIMS) the QPO frequency increases while the radii appear to stay at similar values initially (with large errors). The further increase in the QPO frequency is accompanied by a decrease in the radius. In J1753 (hard state), the QPO frequency does not show a strong dependence on the radius; the radius and the QPO frequency do not show strong changes in our data. In GX-339, the QPO frequency increases at similar radii during the LHS. By the first observation in the HIMS, the QPO frequency and the radius increase by a large factor (from 0.4 Hz to 1.25 Hz and from 110 km to 285 km, respectively). The sharp rise in the QPO frequency continues, as shown by \cite{motta339}, but the radius decreases during the rest of the observations in the HIMS. The decrease in radius is sharp only between the first two HIMS observations (285 km to 92 km) and the decrease is not that dramatic for the rest of the HIMS observations, as seen in Figure \ref{fig:rad-flux}. Hence, it can be seen that in these sources the QPO frequency changes both in association, and in the absence, of strong changes in the color radius.\\
\section{Discussion}\label{discussion}
\noindent BHBs are generally found in the hard states at the start of the outburst. As the outburst progresses, most sources show increasing disk temperature and the X-ray variability becomes weaker. These properties are common across sources despite differences in their physical conditions such as the size (binary separation), orbital period, length of the outburst or states displayed during outburst. Hence, it should be examined what common physical process(es) drive(s) the spectral and temporal phenomena we observe in these systems. \\
\\
In this work we study the outbursts of three BHBs MAXI J1659--152, SWIFT J1753.5--0127 and GX 339--4. These systems are different in terms of their size (binary separation) and the states displayed during outburst. GX-339 has a long orbital period, but the evolution along different states is similar to J1659. Both systems exhibit typical outburst behavior, except that J1659 did not make a transition to the HSS. J1659 and J1753 have short orbital periods, but their evolution along the outburst is dramatically different. J1753 is a quasi persistent system, in outburst since 2005. It remained in the hard state during the rise and decay of the outburst in 2005. It has been suggested that the size of the system may not determine the number of states exhibited, as the same source shows varied behavior in different outbursts \citep[see e.g.,][]{belloni1550}. The size of the system and hence the size of the accretion flow may not be the main factor driving the differences in the outbursts of these sources. \\
\\
Our observations cover the LHS, HIMS and SIMS in GX-339, HIMS and SIMS in J1659 and the hard state of J1753 (see Table \ref{tab:det}). The disk temperature in J1659 and GX-339 spans a large range of values, but in J1753 the disk is cooler and spans a small temperature range. The disk color radii show a varied behavior in the different spectral states, suggesting that the size of the disk varies by different factors in each source. Irrespective of this, the variability components are observed at mostly similar frequencies and amplitudes. We investigate the correlations of each temporal component with the spectral properties within and across different spectral states.\\
\\
We compare our observations with the predictions of the models which attempt to jointly explain the broad-band variability and type-C QPO, discussed in Section \ref{intro}. The models of \cite{zycki-var-model-2003} and \cite{cabanac2010} provide general predictions that the overall strength of variability decreases and the frequency of the components increase as the source evolves to softer states. Also the models predict that strength of variability is higher at high energy. Our results are in agreement with these predictions for the overall behavior along the outburst of the break, the hump and the type-C QPO components; the origin and evolution of lfn (or a similar component) is not discussed in these models. As we study the detailed evolution of each of the variability components, predictions more specific to the origin and evolution of individual variability components are necessary. As the hot flow and Lense-Thirring precession model \citep{ingram-lt-2011} integrates detailed predictions for the different variability components, we discuss below in detail our results in the context of this model.

\subsection{The QPO}
\noindent We first discuss the behavior of the characteristics of the QPO, which shows stronger correlations with spectral properties than the rest of the variability components. In most BHBs, the QPO typically exhibits an increase in frequency and a decrease in strength as the flux and disk temperature increase i.e., the disk contribution increases \citep[e.g.,][]{bhb-mcclintock-2006}. In our data and from earlier reports, we observe similar behavior in J1659 and GX-339 during the HIMS. In GX-339, which is observed also in the LHS, the QPO shows a behavior different than in the HIMS; the frequency increases and the strength decreases, but not in correlation with the disk temperature. In the LHS, the disk temperature does not show a gradual increase as seen in the HIMS. J1753 is observed in the hard state with most observations during the decay of the outburst. The QPO frequency decreases and the amplitude increases, as the disk temperature possibly also decreases. Hence, the pattern is similar to the other sources in the HIMS, but traced  during the decay of the outburst rather than during the rise.  \\
\\
The Lense-Thirring precession model \citep{ingram-lt-2011} predicts that the frequency increases and the \textit{rms} decreases, as the disk moves to smaller radii and the source evolves to softer states. In the case of J1659 and GX-339 in the HIMS, the behavior of the frequency and \textit{rms} discussed above is as predicted by the model. The \textit{rms} of the QPO is weaker in the soft band than the hard band as expected, as the QPO is suggested to originate in the hot flow \citep{sobol-qpo-spectra2006, qpo-lt-axelsson2014}. A lower \textit{rms} in the soft band has been attributed to increased contribution (of non-modulated photons) from the disk in the soft band \citep{ingram2012-frame-dragging}. During the LHS in GX-339, a weaker dependence of the frequency and the \textit{rms} on the increasing disk temperature is observed than in the HIMS.\\
\\
The QPO frequency evolution is predicted by the model to be tied to the changes in the disk truncation radius. The predicted anti-correlation of the frequency with the radius is seen during the HIMS in J1659 but is not clearly seen in GX-339 and J1753 \citep[however, see][]{chiang2010}. Changes in the frequency of the QPO are observed with or without changes in the radius. This might indicate that the behavior of the QPO can only be partly explained by the Lense-Thirring precession model, but could also signify that the inner disk radius is not always reliably estimated by our spectral model. This could be the case in the hard state for the observations where the Comptonization/hard component dominates, and the \texttt{diskbb} model used herein may not properly account for the interaction of the disk with the hard component \citep{shimura1995, diskbb-merloni-2000, spec-salvesen-2013, reynolds-swift-bhs}.

\subsection{The hump component}
\noindent The hump component, which is the broad band noise underlying the QPO, generally exhibits behavior similar to the QPO. The correlation of the frequency with the flux and  the disk temperature seen in J1659 is typical of the HIMS. During the LHS in GX-339 and in the hard state in J1753, the hump frequency does not exhibit a clear correlation with the flux or the disk temperature, at variance with what we observe in the HIMS. The \textit{rms} is weaker in the soft band than the hard band, with the difference being more pronounced in J1753. In the hard band, the \textit{rms} does not decrease with increasing disk temperature and is at similar values over a broad range of disk temperatures in the three sources. This shows that the rise in disk contribution does not strongly affect the strength of this component at harder energies in the hard states. \\
\\
The origin of the hump component has been suggested to be in the inner part of the hot  flow \citep{ingram-lt-2011}. It was shown in J1659 \citep[HIMS,][]{kalamkar1659-swift} and in J1753 \citep[hard state,][]{kalamkar1753.5}, that the component is not detected below 1 keV and the strength increases with energy. This could explain why the increased disk emission does not show an observable effect in the hard band. The drop in \textit{rms} in the soft band could be due to dilution from the disk emission as suggested by \cite{kalamkar1753.5} in J1753. The different behavior of the frequency during the LHS and HIMS cannot be understood. 

\subsection{The break component}
\noindent The break component is observed in J1659 (HIMS) and J1753 (hard state), but not detected in GX-339 in our data. As the break frequency has been associated with the viscous time scale at the truncation radius of the disk \citep{ingram-lt-2011}, it is expected to increase in frequency and decrease in amplitude as the source evolves to the soft state (disk becomes hotter and moves inwards). The break component frequency has been reported to increase in J1659 during the rise \citep[in the 2-60 keV band,][]{kalamkar1659-swift}, where we observe that the disk temperature increases and its radius decreases, as predicted by the model. In J1753 the frequency decreases during the decay when the disk is not detected significantly during that period, although there are indications of a possible decrease in the disk temperature from earlier observations. The disk radius does not show large changes during the outburst peak, but a possible decrease during the decay. This is in accordance with the model prediction. \\
\\
As the break component is associated with the truncation radius, the increasing disk emission should strongly affect the \textit{rms} in the soft band. In J1659, there are only two detections in the hard and the soft bands (see Figure \ref{fig:rms-diskt}). The \textit{rms} is at similar values in the respective bands (but with a higher \textit{rms} in the hard band) at similar disk temperatures. It increases during the evolution to the soft state in the 2-60 keV band in the RXTE data, as reported by \cite{kalamkar1659-swift}. These higher \textit{rms} at higher disk temperatures in J1659 cannot be understood in the context of this model. In J1753, the disk temperature (possibly) decreases as the flux decays (and the disk is	 not significantly detected at low fluxes). The suggestive anti-correlation of the \textit{rms} and disk temperature in J1753 is in accordance with the model prediction. The \textit{rms} in the hard and the soft bands increases during the flux decay. Although the \textit{rms} appears to be similar in both the energy bands in our simultaneous detections during the decay, \cite{kalamkar1753.5} showed that the low frequency soft band variability is systematically lower in the soft band than the hard band. 

\subsection{The low frequency noise}
\noindent The lfn is detected in J1659 during the HIMS and the SIMS and in GX-339 during the LHS and the HIMS. It is observed at similar frequencies across all spectral states in both sources. The component has been suggested to arise intrinsically due to fluctuations in the disk \citep[in J1659][]{kalamkar1659-swift, wenfei1659}. The lack of frequency evolution can be understood if the component arises due to fluctuations in the disk at a radius larger than the disk truncation radius i.e., it is not associated with a `moving' disk radius. As the fluctuations propagate to smaller radii (possibly also into the hot flow) and modulate the emission, the strength of this component is expected to be sensitive to the changes in the size of the accretion disk and the disk temperature (particularly in the soft band). In GX-339, the \textit{rms} drops during the LHS and reaches values similar to HIMS in J1659 as the disk temperature increases. During the HIMS in J1659, when the disk temperature strongly increases, the \textit{rms} is observed at similar strengths (with some scatter) over a large range of disk temperatures. Also, there are more detections in the soft band than the hard band, with stronger \textit{rms} in the soft band. It can be clearly seen that the strength of the lfn in the LHS and the HIMS is different. However, it is unclear why the fluctuations giving rise to this component are affected by increasing disk contribution differently in the LHS and the HIMS.\\
\\
In J1753, the lfn is not detected. An earlier report by \cite{kalamkar1753.5} suggested that during the outburst peak, the low frequency variability ($<$ 0.4 Hz) is weaker in the soft band than the hard band due to stronger disk emission (although the disk temperature is low  at $\sim$ 0.2 keV).  However, the detection of lfn in J1659 at high disk temperatures (up to $\sim$ 0.55 keV) makes the dilution due to strong disk emission an unlikely explanation for the non detection of the lfn in J1753. One possibility is that due to the cool disk, the lfn is stronger at soft energies below the XRT limit of 0.5 keV. Hence, the disk emission affects the lfn in different manner in different sources.\\
\\  
Despite this inconsistent behavior of the lfn, it could play an important role in determining if the disk is truncated in the hard state, assuming its origin intrinsic to the disk. In GX-339, it is detected in the LHS even when the disk is not detected significantly. If the lfn has a soft rms spectrum (even if the disk contribution is not statistically significant in the energy spectrum), this would provide support to the earlier suggestions \citep[e.g.][]{reis-bhb-lhs, reynolds-swift-bhs} that the disk may not be truncated far from the black hole in the LHS. If the disk is truncated and the lfn arises in the hot flow in the LHS, it is expected to exhibit a hard rms spectrum similar to other higher frequency components \citep[see e.g.,][]{kalamkar1659-swift}.
\section{Summary}
\noindent This work highlights the importance of correlated spectral and timing studies in the \textit{Swift} bandpass of 0.5-10 keV. The disk temperature, its radius and the effects of disk emission on the different variability components, arising in both the components of the accretion flow, can be studied simultaneously. We conclude that the evolution of the spectral and timing components is complex and cannot be fully explained by any single correlation; there are marked differences in the correlations in different spectral states within a source and amongst different sources. The spectral-timing correlations in J1659 and GX-339 are similar during the HIMS, but different in GX-339 during the LHS and the hard state of J1753. The changes in the disk temperature, its contribution to emission and possibly the size of the disk itself do not affect the different variability components in the same manner. The models discussed in the previous section cannot yet explain all the observed correlations and why these correlations are different in different states of the outburst. These models should also take into account the variability originating in the disk. This could provide a framework to explain the behavior of the lfn. Our results provide a framework to further develop these models (or develop new ones) which can take into account the spectral and temporal behaviour of emission not only in the hard band, but also in the soft band. Also taking into account the uncertainties associated with spectral parameters, more accurate estimates are necessary to understand the nature of the correlations observed in the hard state. Similar studies with \textit{XMM-Newton} and perhaps LOFT in the future, which have the sensitivity to observe these sources at lower fluxes than \textit{Swift}, has a strong potential to resolve these issues. \\
\\
\section*{Acknowledgements}
\noindent This research has made use of data obtained from the High Energy Astrophysics Science Archive Research Center (HEASARC), provided by NASA's Goddard Space Flight Center, and also made use of NASA's Astrophysics Data System. DA acknowledges support from the Royal Society.

\bibliographystyle{aa}
\bibliography{bibliography}
\end{document}